\documentclass[aps,prb,twocolumn]{revtex4}

\usepackage{amsmath,amssymb,mathrsfs}
\usepackage[dvips]{graphicx}
\usepackage{psfrag,latexsym}

\def\ea{\emph{et al.}}

\allowdisplaybreaks[4]

\begin{document}
\title{Relaxation vs decoherence: Spin and current dynamics in the 
  anisotropic\\ Kondo model at finite bias and magnetic field}
\author{Mikhail Pletyukhov}
\author{Dirk Schuricht}
\author{Herbert Schoeller}
\affiliation{Institut f\"ur Theoretische Physik A, RWTH Aachen, 
  52056 Aachen, Germany}
\affiliation{JARA-Fundamentals of Future Information Technology}
\date{\today}
\pagestyle{plain}

\begin{abstract}
  Using a nonequilibrium renormalization group method we study the real-time
  evolution of spin and current in the anisotropic Kondo model (both
  antiferromagnetic and ferromagnetic) at finite magnetic field $h_0$ and bias
  voltage $V$. We derive analytic expressions for all times in the
  weak-coupling regime $\max\{V,h_0,1/t\}\gg T_c$ ($T_c=$ strong coupling
  scale). We find that all observables decay both with the spin relaxation and
  decoherence rates $\Gamma_{1/2}$.  Various $V$-dependent logarithmic,
  oscillatory, and power-law contributions are predicted. The low-energy
  cutoff of logarithmic terms is generically identified by the difference of
  transport decay rates. For small times $t\ll \max\{V,h_0\}^{-1}$, we obtain
  universal dynamics for spin and current.
\end{abstract}
\pacs{73.63.Kv}
\maketitle

The real-time dynamics of small strongly interacting quantum systems coupled
to several reservoirs (e.g. quantum dots, quantum impurities, or single
molecules) is a fundamental nonequilibrium problem. The interest in this field
stems from the experimental progress in controlling spin dynamics in quantum
dots \cite{QI_general}, and from the necessity to identify the qualitative
dynamics for error correction schemes \cite{Loss_general}.  The theoretical
description of such situations remains a huge challenge.  Numerical techniques
like time-dependent numerical\cite{TD_NRG_general,Roosen_etal_08} and density
matrix renormalization group methods\cite{TD-DMRG_general}, iterative
path-integral methods\cite{Weiss_etal_08}, and nonequilibrium Monte Carlo
simulations\cite{Schmidt_etal_08} have been developed to describe the
time-evolution.  However, the description of finite bias or the long-time
limit is often difficult.  Analytically very little is known, except for
special models which can be solved exactly \cite{Exact_general}. For this
reason, perturbative renormalization group (RG) methods have been developed
for nonequilibrium problems
\cite{RTRG_general,rosch_paaske_kroha_woelfle_03,Kehrein_general,Hackl_etal_08,
  RTRG-FS_1,RTRG-FS_2,RTRG-FS_3} to obtain results in the regime of weak
coupling between dot and reservoirs. Concerning the time evolution, these
methods have so far been applied to the spin boson model
\cite{RTRG_spin_boson_01,Hackl_etal_08} and to the Kondo model at zero voltage
for special regimes \cite{Lobaskin_Kehrein_05,Hackl_etal_09}.

In this Letter, we will use a recently developed real-time RG method in
frequency-space\cite{RTRG-FS_1,RTRG-FS_2,RTRG-FS_3} to calculate
analytically the full time-evolution of the anisotropic Kondo model at finite
voltage $V$ and magnetic field $h_0$. This method has the particular advantage
that the derived exact hierarchy of RG equations contains the full dynamics of
local observables. In order to solve the RG equations systematically by
expanding in the renormalized couplings, we will consider the weak-coupling
regime where at least one physical low-energy scale is larger than the strong 
coupling scale $T_c$. For the Kondo model at zero temperature we consider
\begin{equation}
\label{eq:cutoff_scale}
\Lambda_t=\max\{V,h_0,1/t\}\gg T_c,
\end{equation}
where $T_c=T_K$ (Kondo temperature) for the antiferromagnetic (AFM) and
$T_c=0$ for the ferromagnetic (FM) Kondo model. In this regime, the couplings
$J\equiv J(\Lambda_t)$ can be chosen as expansion parameters, where
$J(\Lambda)$ denotes the poor man scaling (PMS) solution at scale $\Lambda$
(see \eqref{eq:RG_leading_solution} below).  We find several interesting
results which are proposed to be generic for any quantum dot in the Coulomb
blockade regime where spin/orbital fluctuations dominate: 1. The voltage is an
important energy scale for the dynamics which shows up in oscillatory,
power-law, or logarithmic behavior.  2. In the long-time limit $t\gg
1/\max\{V,h_0\}$, we find generically, i.e.  in all orders of perturbation
theory, that all terms are exponentially decaying with the transport rates
$\Gamma_i$. Furthermore, due to non-Markovian effects, each spin component and
the current contain a sum of \emph{two} terms, one decaying with the
relaxation rate $\Gamma_1$ and the other with the decoherence rate $\Gamma_2$.
3. At finite voltage, resonances $V,h\gg |h-V|\rightarrow 0$ or $V\gg
h\rightarrow 0$ ($h$ is the renormalized magnetic field) are possible in the
weak-coupling regime. In the limit $1/\max\{V,h_0\}\ll t\ll 1/|\delta|$ (with
$\delta\equiv h-V,h$), we find logarithmic terms $\sim (\delta t)\ln|\delta
t|$ for the transverse spin. In contrast to stationary
quantities\cite{rosch_paaske_kroha_woelfle_03,glazman_pustilnik_05,RTRG-FS_2}
the cutoff scale of the logarithmic terms for $\delta\rightarrow 0$ is
generically determined by the \emph{difference} of decay rates
$\Gamma_1-\Gamma_2$. 4. In the short-time limit $t\ll 1/\max\{V,h_0\}$, we
obtain universal dynamics for spin and current.

\begin{figure}[t]
  \centering
  \includegraphics[width=60mm,clip=true]{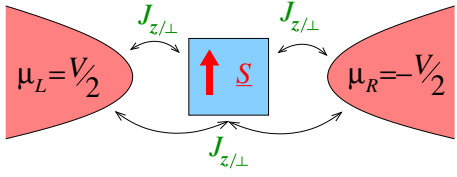}
  \caption{(Color online) A spin-$\frac12$ quantum system coupled via
    exchange couplings $J_{z/\perp}$ to the spins of two reservoirs.}
  \label{fig:kondo}
\end{figure}

\emph{Model.}---The model consists of a single spin-$\frac12$, which is
coupled by longitudinal and transverse exchange couplings $J^{(0)}_{z/\perp}$
to the spins of two noninteracting reservoirs, see Fig.~\ref{fig:kondo} for a
sketch of the system. Experimental realizations of the model are provided by 
quantum dots \cite{goldhaber} and molecular magnets\cite{SMM_theory}.  
The Hamiltonian reads
$H=H_{res}+H_D+H_{ex}$, where $H_{res}=\sum_{\alpha\sigma k}\epsilon_{\alpha
  k}a^\dagger_{\alpha\sigma k}a_{\alpha\sigma k}$ describes two noninteracting
reservoirs labeled by $\alpha=L,R$ ($\sigma=\downarrow,\uparrow$ denotes the
spin and $k$ is the state index), $H_D=h_0 \hat{S}_z$ is the Hamiltonian of
the local spin with Zeeman splitting, and
\begin{equation}
\label{kondo}
H_{ex}=J^{(0)}_\perp\,(\hat{S}_x\hat{s}_x+\hat{S}_y\hat{s}_y)
+J^{(0)}_z\,\hat{S}_z\hat{s}_z
\end{equation}
denotes the coupling\cite{endnote_1}. We use the notation
$\underline{\hat{s}}=\frac12 \sum_{\alpha\sigma k\alpha'\sigma' k'}
a^\dagger_{\alpha\sigma k}
\,\underline{\sigma}_{\sigma\sigma'}\,a_{\alpha'\sigma' k'}$
($\underline{\sigma}$ are the Pauli matrices).  The reservoirs are kept at
different chemical potentials $\mu_{L/R}=\pm V/2$ and are assumed to have a
flat density of states in the band of width $2D$. Furthermore, we consider the
most interesting case of zero temperature.

\emph{Method.}---To study the dynamics of the local spin
$\underline{S}(t)=\langle \underline{\hat{S}}\rangle(t)$ and the charge
current $I(t)=\langle \hat{I}\rangle(t)$, we switch on the coupling $H_{ex}$
suddenly at the initial time $t=0$. It means that we prepare the initial
density matrix in the product form $\rho(0)=\rho_D^{(0)}\rho_L\rho_R$, where
$\rho_D^{(0)}$ is an arbitrary density matrix for the local spin, and
$\rho_{L/R}$ are grand canonical distributions for the left and right
reservoirs. For $t>0$, we calculate the dynamics from a kinetic equation for
the reduced density matrix $\rho_D(t)=\text{Tr}_{res}\rho(t)$. Following
Ref.~\onlinecite{RTRG-FS_1}, one obtains in Laplace space and in Liouvillian
notation $\rho_D(z)=i/(z-L_D^{eff}(z))\,\rho^{(0)}_D$, where $L_D^{eff}(z)$ is
an effective dot Liouvillian.  The spin dynamics follows from
$\underline{S}(t)=\text{Tr}_D\,\hat{\underline{S}}\,\rho_D(t)$, with
$\rho_D(t)=\frac{1}{2\pi}\int dz \,e^{-izt}\rho_D(z)$. The current in Laplace
space is given by $I(z)=-i\text{Tr}_D \Sigma_I(z)\rho_D(z)$, where
$\Sigma_I(z)$ denotes the current kernel. $L_D^{eff}(z)$ and $\Sigma_I(z)$
have been calculated in Ref.~\onlinecite{RTRG-FS_2} for an arbitrary quantum
dot in the Coulomb blockade regime with explicit formulas for the anisotropic
Kondo model. The result is given by a systematic expansion in the
dot-reservoir couplings from PMS cut off at scale
$\Lambda_z=\max\{V,h_0,|z|\}\gg T_c$.  The PMS equations are well-known
\cite{solyom_zawadowski_74} and given by
$(d/d\Lambda)J_{z/\perp}(\Lambda)=-2J_\perp(\Lambda)
\,J_{\perp/z}(\Lambda)/\Lambda$, with the solution
\begin{equation}
\label{eq:RG_leading_solution}
J_z(\Lambda)=c\, \frac{1+\left(\frac{T_K}{\Lambda}\right)^{4c}} 
   {1-\left(\frac{T_K}{\Lambda}\right)^{4c}},\;
J_\perp(\Lambda)=2c\, \frac{\left(\frac{T_K}{\Lambda}\right)^{2c}} 
   {1-\left(\frac{T_K}{\Lambda}\right)^{4c}},
\end{equation}
where $c^2=J_z^2-J_\perp^2$ and $T_K$ are two invariants.  In the isotropic
case one obtains $J_{z/\perp}(\Lambda)=1/(2\ln(\Lambda/T_K))$.  For a study of
the time dynamics it appears to be more convenient to expand in the couplings
cut off at the time-dependent scale $\Lambda_t$ defined in
\eqref{eq:cutoff_scale}. We have proven that this expansion is well-defined in
the weak-coupling regime $\Lambda_t\gg T_c$ and equivalent to the expansion in
$J(\Lambda_z)$.

In conventional perturbation theory one 
approximates $L_D^{eff}(z)\approx L_D^{eff}(z=0)$ (Markov approximation) and
considers only the terms up to second order in the bare exchange couplings
$J_{z/\perp}^{(0)}$ (Born approximation). This approach fails to describe the
current dynamics as well as yields simple exponential decay for the 
longitudinal (transverse) spin with the spin relaxation (decoherence) rate 
$\Gamma_1$ ($\Gamma_2$).  In contrast, in this Letter we include non-Markovian 
terms and use a systematic perturbative approach in the renormalized couplings
$J_{z/\perp}$. 
In particular, replacing the effective Liouvillian in the resolvent
$1/(z-L_D^{eff}(z))$ by any of its nonzero eigenvalues $\lambda_p(z)$
evaluated up to $O(J^2)$, we obtain an expression of the generic form
\begin{equation}
\tilde{Z} \biggl[ z-\widetilde{\Delta}_p+{\sum_b}c_b J_b^2\,
(z-z_b)\ln \frac{\Lambda_t}{ -i(z-z_b)}\biggr]^{-1}.
\label{eq:lambda_generic}
\end{equation}
The pole of this resolvent is denoted by $z_p=\lambda_p(z_p)$.  For the Kondo
model, there are three nonzero poles ($p=1,\pm$) at $z_1=-i\Gamma_1$ and
$z_\pm=\pm h-i\Gamma_2$, where $\Gamma_1=\pi J_\perp^2(h+\max\{V,h\})$ and
$\Gamma_2=(\pi/2)J_z^2 V+\Gamma_1/2$. The real parts of these poles up to
$O(J)$ determine the quantities $\widetilde{\Delta}_p$,
i.e.  $\widetilde{\Delta}_\pm =\pm(1-J_z+J_z^{(0)})h_0$ and
$\widetilde{\Delta}_1=0$.  Besides the appearance of renormalized quantities,
there are two new important contributions in (\ref{eq:lambda_generic}).
Firstly, the factor $\tilde{Z}=1-2(J_z-J_z^{(0)})$ occurs, which contains
$J_z\equiv J_z(\Lambda_t)$ and thereby determines the universal
non-exponential short-time behavior. Secondly, the logarithmic part is also of
non-Markovian form. It induces branch cuts in the complex plane with branch
points $z_b$ located at $z_b=\pm h+nV-i\Gamma_2$ (for $p=1$) and
$z_b=nV-i\Gamma_1$ or $z_b=\pm h+nV-i\Gamma_2$ (for $p=\pm$), with $n=0,\pm 1$
($c_b$ and $J_b\equiv J_{z/\perp}$ are the certain coefficients and
couplings). The branch cuts lead to exponential behavior on the scale $z_b$,
\emph{not} $z_p$, i.e. with oscillation frequencies involving the voltage as 
well as unexpected decay rates. These exponential terms are multiplied by 
additional
power-law or logarithmic functions involving the scale $z_b-z_p$.  Therefore,
all logarithmic terms appearing in the long-time evolution are cut off by the
\emph{difference} of decay rates $\Gamma_{pp'}=\Gamma_p-\Gamma_{p'}$, a
feature which is generic for all quantities entering the time evolution
(cf. Ref.~\onlinecite{RTRG-FS_2}).

\emph{Results.}---Considering all poles and branch cuts of the resolvents
(\ref{eq:lambda_generic}) and closing the integration contour of the inverse
Laplace transform in the lower half plane of $z$, we obtain after some lengthy
algebra the following final result for the spin and current dynamics in closed
form (we set $e=\hbar=1$ and assume $V,h>0$)
\begin{widetext}
\begin{eqnarray}
\label{eq:spin_longitudinal}
S_z(t)&=&S_z^{(0)}\,\tilde{Z} e^{-\Gamma_1 t}+S_z^{st}(1-e^{-\Gamma_1 t})
-S_z^{(0)}J_\perp^2\,\text{Re}\!
\left[S^{h+i\Gamma_{12}}_{z_1 z_+}+S^{h-V+i\Gamma_{12}}_{z_1,z_+-V}+
(V\rightarrow -V)\right],\\
\label{eq:spin_transverse}
S_+(t)&=&S_+^{(0)}\,e^{-\Gamma_2 t}e^{iht}
\left[\tilde{Z}-J_z^2\ln(\Lambda_t t)
\right] - \frac{1}{2}\,S_+^{(0)}\left[J_z^2 S^{V}_{z_-,z_-+V}+
J_\perp^2 (S^{h-i\Gamma_{12}}_{z_-z_1}+
S^{h-V-i\Gamma_{12}}_{z_-,z_1-V}) + (V\rightarrow -V)\right],\\
\label{eq:current}       
I(t)&=&\frac{\pi}{4} (J_z^2+2J_\perp^2)V
+\pi J_\perp^2\min\{V,h\}S_z^{st}(1-e^{-\Gamma_1 t})
-\frac{1}{t}\,S_z^{(0)}J_\perp^2\,\text{Re}\!
\left[I^{h-V+i\Gamma_{12}}_{z_1,z_+-V}-(V\rightarrow -V)\right],
\end{eqnarray}
\end{widetext}
where $S_+=S_x+iS_y$ and $\Gamma_{12}=\Gamma_1-\Gamma_2$. The initial and the
stationary magnetizations are denoted by
$\underline{S}^{(0)}=\underline{S}(0)$ and $S_z^{st}=-h/(h+\max\{V,h\})$, and
we have introduced the auxiliary functions $S^\alpha_{zz'}(t)=e^{-izt}\ln
\frac{\Lambda_t}{i\alpha}+ e^{-iz't} F_2 (\alpha t)$ and
$I^\alpha_{zz'}(t)=e^{-izt}i\alpha t \ln
\frac{\Lambda_t}{i\alpha}+e^{-iz't}F_1 (\alpha t)$, with $F_k
(x)\!=\!(-1)^{k-1} \!\int_0^\infty dy\,y e^{-y}/(y+ix)^k$.

\begin{figure}[t]
  \centering
  \includegraphics[width=80mm,clip=true]{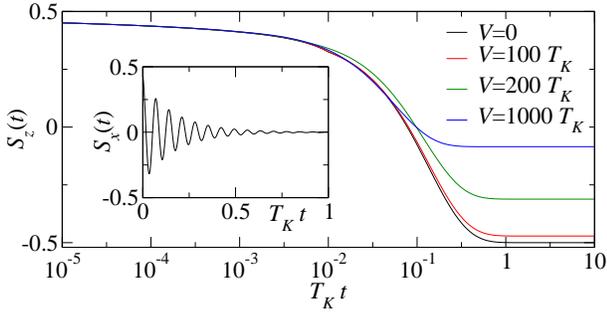}
  \caption{(Color online) $S_z(t)$ in the isotropic Kondo model for 
    $h_0=100\,T_K$ and various values of the applied voltage $V$, with
    $S_z(0)=1/2$. Inset: $S_x(t)$ for $V=h_0=100\,T_K$ and $S_x(0)=1/2$,
    $S_y(0)=0$.}
  \label{fig:magnetization}
\end{figure}

\begin{figure}[t]
  \centering
  \includegraphics[width=80mm,clip=true]{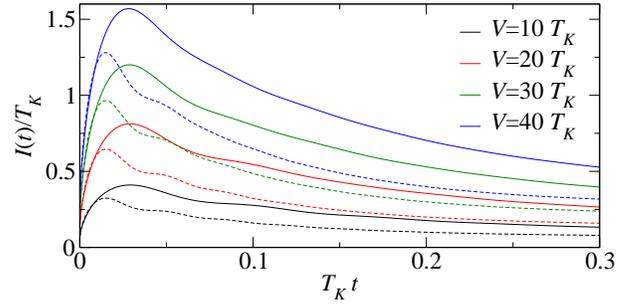}
  \caption{(Color online) Current $I(t)$ in the isotropic Kondo model for 
    $h_0=100\,T_K$ (solid lines) and $h_0=200\,T_K$ (dashed lines) and 
    various values of the applied voltage $V$, with $S_z(0)=1/2$. We observe
    oscillations at times $T_Kt\sim 0.05$; the stationary current is
    reached for $T_Kt\ge 0.5$.}
  \label{fig:current}
\end{figure}

Equations \eqref{eq:spin_longitudinal}-\eqref{eq:current} provide the time
dynamics up to $O(J^2)$ in all crossover regimes. We stress that all
observables decay both with the relaxation and decoherence rates. Correction
terms $\sim O(J^2)$ oscillate with frequencies $h\pm V$, and, additionally,
with $h$ ($nV;\, n=0,\pm$) for the longitudinal (transverse) spin.  From an
experimental point of view these oscillations along with the occurrence of
both decay rates are our most important results.  Examples for the time
evolution in the isotropic case $J_z=J_\perp$ are shown in
Figs.~\ref{fig:magnetization} and~\ref{fig:current}. We now illustrate the
results in different time regimes.

\begin{figure}[t]
  \centering
  \includegraphics[width=80mm,clip=true]{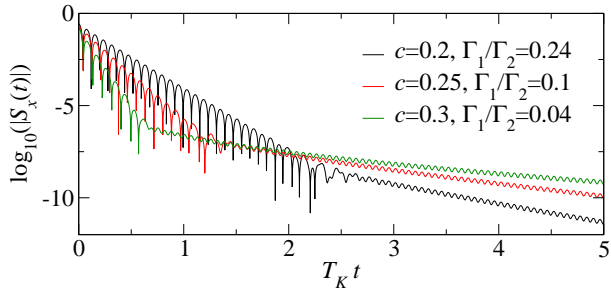}
  \caption{(Color online) $\log_{10}(|S_x(t)|)$ in the anisotropic Kondo 
    model for $V=2h_0=100\,T_K$ and various values of the anisotropy
    $c^2=(J_z^{(0)})^2-(J_\perp^{(0)})^2$, with $S_x(0)=1/2$, $S_y(0)=0$.  The
    dips have their origin in the oscillations of $S_x(t)$.}
  \label{fig:anis}
\end{figure}

\emph{Long-time limit, off-resonance.}---For large times
$t\gg\max\{V,h_0\}^{-1}$, we have $\Lambda_t=\Lambda_c\equiv\max\{V,h_0\}$ and
$J_{z/\perp}\equiv J_{z/\perp}(\Lambda_c)$.  In the off-resonance case
$V,h,|h\pm V|\sim\Lambda_c$, we get typical power-law behavior from the
asymptotic expansions $F_1(x)\approx -i/x$ and $F_2(x)\approx 1/x^2$, for
$|x|\gg 1$. Besides the exponential decay, this gives additional factors $\sim
1/(\alpha t)^k$, with $k=1$ for the current and $k=2$ for the spin. For the
transverse spin we obtain, for example,
\begin{eqnarray}
\label{eq:spin_transverse_off_resonance}
S_+(t)&\approx&S_+^{(0)}e^{-\Gamma_2 t}e^{iht}
\left[Z-J_z^2 \frac{\cos(Vt)}{V^2 t^2}\right]\\\nonumber
&&\hspace{-1.5cm}-S_+^{(0)} \frac{J_\perp^2}{2}
e^{-\Gamma_1 t}\left[ \frac{2}{h^2 t^2} + \frac{e^{iVt}}{(h-V)^2 t^2}
+ \frac{e^{-iVt}}{(h+V)^2 t^2}\right],
\end{eqnarray}
with $Z=\tilde{Z}-J_z^2\ln(\Lambda_c t) +i \frac{\pi}{2}
\{J_z^2+J_\perp^2[2-\Theta(V-h)]\}$.  To identify the different time scales
from the exponential decay explicitly, situations where $\Gamma_1$ and
$\Gamma_2$ differ significantly are of particular interest.  For the
anisotropic Kondo model at finite voltage this can be easily achieved since
$\Gamma_2/\Gamma_1=1/2+O((J_z/J_\perp)^2)$. Thus, for $J_z\gg J_\perp$ we
obtain $\Gamma_2\gg\Gamma_1$. This is typically the case for single molecular
magnets, where the transverse exchange coupling is generated by quantum
tunneling of magnetization\cite{SMM_theory} (e.g. in Fe$_4$ the ratio is
$J^{(0)}_\perp/J^{(0)}_z\sim 10^{-5}$). Fig.~\ref{fig:anis} shows the clear
separation of time scales for $S_x(t)$.  The time scale at which the change of
the leading behavior happens is approximately given by $t\sim 2|\ln
J_\perp|/\Gamma_{21}$.  At this time we also observe the change in the
oscillation frequency from $h$ to $V$.

\emph{Long-time limit, on-resonance.}---For large times $t\gg\Lambda_c^{-1}$
and close to resonances, where $|\delta|\ll\Lambda_c$, with $\delta=h-V$ or
$\delta=h$, we can enter the interesting time regime $|\delta t|\ll 1$. In
this case, we obtain logarithmic contributions from $F_k (\alpha t)$, with
$|\alpha t|\ll 1$ and $\alpha=\delta-i\Gamma_{12}$. For the longitudinal spin
and the current, the logarithmic terms $\sim\ln|\delta t|$ cancel out.
In turn, for the transverse spin we obtain the logarithmic contributions
\begin{equation}
\label{eq:spin_transverse_on_resonance}
S_+(t)\rightarrow -a\,e^{i\pi/2}\,S_+^{(0)}\,
J_\perp^2 \,\delta t\ln|(\delta-i\Gamma_{12})t|,
\end{equation}
with $a=1/2,1$ for $\delta=h-V,h$. As a consequence, the spin perpendicular to
the plane defined by the $z-$axis and the initial direction obtains a
logarithmic enhancement at resonance $\delta\rightarrow 0$ after taking the
derivative $\frac{d}{d\delta}$. The cutoff of the logarithm at $\delta=0$ is
determined by the difference $\Gamma_{12}$ of the rates. Especially for
$\delta=h\ll V$ in the isotropic model the cutoff scale
$\Gamma_{12}=\frac{\pi}{2}J^2 h$ becomes zero close to resonance. In this case
the perturbative approach breaks down for exponentially small values of $ht$
where $J\ln|ht|\sim O(1)$. This regime is beyond our approach and requires a
nonperturbative treatment.

\emph{Short-time limit.}---For short times\cite{ultra_short} $ D^{-1} \ll t\ll
\Lambda_c^{-1}$ we get $\Lambda_t=1/t$ and $J_{z/\perp}\equiv
J_{z/\perp}(1/t)\equiv J_{z/\perp}^t$.  The terms containing
$(S/I)^\alpha_{zz'}$ are negligible and we obtain from the first terms of
\eqref{eq:spin_longitudinal}-\eqref{eq:current} the universal result
\begin{eqnarray}
\label{eq:spin_universal}
\underline{S}(t)&=&\tilde{Z} \,\underline{S}^{(0)}\,=\,
(1-2J_z^t+2J_z^{(0)})\,\underline{S}^{(0)},\\
\label{eq:current_universal}
G(t)&=&\frac{I(t)}{V}\,=\,\frac{e^2}{h} \frac{\pi^2}{2}
\left[(J_z^t)^2+2(J_\perp^t)^2\right].
\end{eqnarray}
For $J_z^{(0)}\gg J_\perp^{(0)}$ and $|(T_Kt)^{4c}|\ll 1$ (AFM case), we get
\begin{eqnarray}
\label{S_AFM_universal}
\underline{S}(t)&=&\underline{S}(0)\{1+2J_z^{(0)}-2c-4c(T_K t)^{4c}\},\\
\label{G_AFM_universal}
G(t)&=&\frac{e^2}{h}\left[\frac{\pi^2}{2}c^2\,+\,6\pi^2 c^2 (T_K t)^{4c}
\right],
\end{eqnarray}
with $c=\sqrt{(J_z^{(0)})^2-(J_\perp^{(0)})^2}$. For $|(T_Kt)^{4c}|\gg 1$ (FM
case), we have to replace $c\rightarrow -c$.  As a result, universal
power laws are predicted in the short-time limit on the scale of the Kondo
temperature. We note that for the FM Kondo model with $V=h_0=0$ a power law
has also been found for $S_z(t)$ in Ref.~\onlinecite{Hackl_etal_09}. Here, we
have found a complete analytic expression in terms of the PMS solution,
together with results for the AFM model, the transverse spin, and the current.
Furthermore, we identified the validity range for finite $\Lambda_c$ by $t\ll
\Lambda_c^{-1}$.  In the isotropic case ($c=0$), we obtain $J^t=-1/(2\ln(T_K
t))$ leading to $\underline{S}(t)=\underline{S}(0)(1+2J_z^{(0)}+1/\ln(T_K t))$
and $G(t)=(e^2/h)3\pi^2/(8\ln^2(T_K t))$ for both AFM and FM models. In this
case, universal logarithmic terms occur, which have also been found for
$S_z(t)$ in Refs.~\onlinecite{Roosen_etal_08,Hackl_etal_09}.

\emph{Summary.}---We found complex dynamics for a spin coupled to fermionic
reservoirs held at finite bias and for the current across it.  The results go
beyond Markovian theories and, in a well-controlled weak-coupling regime, are
presented in closed analytic form covering all time regimes. Due to the
generic form of Eq.(\ref{eq:lambda_generic}), we propose our main conclusions
to be generic for quantum dots in the Coulomb blockade regime. From
Ref.~\onlinecite{RTRG-FS_2} it follows that the branch points of the
logarithmic terms are generically given by $z_b=z_p+nV$, where $z_p$ is any
pole of the leading order exponential decay, and $n=0,\pm 1,\pm 2,\dots$.
Therefore, we find in all orders of perturbation theory that all terms are
exponentially decaying, all transport rates occur in principle in the decay of
each observable, and the voltage appears in the oscillation frequencies. We
have shown this explicitly for the Kondo model and revealed that the bias
voltage is an important energy scale for the time evolution showing up in
unexpected oscillation frequencies and in various power-law and logarithmic
contributions.

This work was supported by the DFG-FG 723 and 912.


\begin{thebibliography}{99}
\bibitem{endnote_1}
We assume the exchange couplings to be independent of the 
reservoir index. Other cases can be equally well treated.
\bibitem{ultra_short} At ultra-short times $t \ll D^{-1}$ an expansion in
$J_{z/\perp}^{(0)}$ and $D t$ is applicable. The scale $T_K$ is
invisible in this regime.
\end{thebibliography}
\end{document}